\documentclass[11pt, bibliography=totocnumbered,headsepline,abstract=true]{scrartcl}
\usepackage[paper=a4paper, left=25mm, right=25mm, top=25mm, bottom=30mm,marginparwidth=2.75cm]{geometry}

\usepackage{amsmath, amssymb, graphicx, url,hyperref}

\usepackage[usenames,dvipsnames]{color}
\usepackage[english,capitalize]{cleveref} 

\usepackage{subcaption}
\usepackage{acronym}
\usepackage{soul}
\usepackage[english]{babel}
\usepackage{microtype}

\newcommand{\threesubsection}[1]{\subsection{#1}} 

\newcommand\blankfootnote[1]{%
  \begingroup
  \renewcommand\thefootnote{}\footnotetext{#1}%
  \addtocounter{footnote}{-1}%
  \endgroup
}

\usepackage[backend=biber, style=science, citestyle=numeric-comp, sorting=none, maxnames=3, sortcites=true, date=year, isbn=false, doi=true, url=false, eprint=false]{biblatex}
\begin{document}

\title{ {Transversal Halide Motion Intensifies Band-To-Band Transitions in Halide Perovskites}}

\author{
Christian Gehrmann$^+$,
Sebasti\'{a}n Caicedo-D\'{a}vila$^+$,
Xiangzhou Zhu,
\\ and David A. Egger$^*$
}

\date{} 

\maketitle

\begin{center}
Department of Physics, Technical University of Munich, Garching, Germany
\end{center}

\blankfootnote{$^+$These authors contributed equally}
\blankfootnote{$^*$Email Address: david.egger@tum.de}


\begin{abstract}
Despite their puzzling vibrational characteristics that include strong signatures of anharmonicity and thermal disorder already around room temperature, halide perovskites exhibit favorable optoelectronic properties for applications in photovoltaics and beyond.
Whether these vibrational properties are advantageous or detrimental to their optoelectronic properties remains, however, an important open question.
Here, this issue is addressed by investigation of the  {finite-temperature optoelectronic properties} in the prototypical cubic CsPbBr$_3$, using first-principles molecular dynamics based on density-functional theory.
It is shown that the dynamic flexibility associated with halide perovskites enables the so-called \textit{transversality}, which manifests as a preference for large halide displacements perpendicular to the Pb-Br-Pb bonding axis.
We find that transversality is concurrent with vibrational anharmonicity and  {leads to a rapid rise in the joint density of states}, which is favorable for photovoltaics since this implies sharp optical absorption profiles.
These findings are contrasted to the case of PbTe, a material that shares several key properties with CsPbBr$_3$ but cannot exhibit any transversality and, hence, is found to exhibit much wider band-edge distributions.
We conclude that the dynamic structural flexibility in halide perovskites and their unusual vibrational characteristics might not just be a mere coincidence, but play active roles in establishing their favorable optoelectronic properties.
\end{abstract}

\pagebreak


Halide perovskites (HaPs) are semiconducting materials with an enormous potential for various technological applications, perhaps most notably for photovoltaics.\cite{Snaith2013,Green2014,Stranks2015,Correa-Baena2017} 
Significant interests in these systems are motivated by their favorable optoelectronic properties and the simultaneous availability of low-cost synthesis and fabrication routes.\cite{Brenner2016,Stoumpos2016,Li2017b,Nayak2019} 
Among their many outstanding physical and chemical characteristics, the structural dynamical properties of HaPs are especially intriguing.\cite{Frost2016a,Egger2016,Egger2018}
In particular, significant vibrational anharmonicities have been detected in these and related systems already at room temperature.\cite{Whalley2016,Beecher2016,Yaffe2017,Carignano2017,Marronnier2017,Gold-Parker2018,Marronnier2018,Zhu2019,Sharma2020,Klarbring2020} 
The unusual lattice vibrational properties of HaPs have already been shown to influence some key optoelectronic quantities including the band-gap energy,\cite{Lanigan-Atkins2021} 
charge-carrier mobilities, \cite{Mayers2018,Lacroix2020,Schilcher2021} 
defect energetics, \cite{Cohen2019,Wang2019b,Li2019b,Kumar2020,Zhang2020,Chu2020} 
the Urbach energy, \cite{Wu2019,Gehrmann2019} 
carrier recombination rates and ion migration barriers,\cite{Bischak2017,Kirchartz2018,Munson2018,Munson2019} 
 {It is an important open question to what extent the peculiar lattice vibrational properties of HaPs influence their optoelectronic properties.}\cite{Munson2019}
Since vibrational anharmonicities are known to occur in other optoelectronic materials,\cite{Delaire2011,Li2014a,Li2014b,Lunghi2017,Svane2017,Kapil2019,Sanni2020,Asher2020,Brenner2020,Menahem2021} 
addressing this question is important because it will support the design of alternative material systems with similarly favorable or improved optoelectronic properties.

In this work, we perform first-principles molecular dynamics (MD) simulations to investigate the origins and consequences of the unusual lattice vibrational properties of HaPs.
 {Specifically, the goal of our work is to investigate the connections between the lattice vibrations and the optical absorption profile of HaPs.}
In characterizing the vibrational features of the prototypical cubic CsPbBr$_3$ we highlight a distinctive property -- which we call \textit{transversality} --  that showcases the high degree of \textit{dynamic structural flexibility in HaPs}.
Transversality is shown to derive naturally from the octahedral arrangement of halide ions in the perovskite lattice and to coincide with large vibrational displacements and anharmonicities in the system.
Importantly, the dynamic structural flexibility of HaPs as expressed in transversality is found to have a strong effect on disorder correlations that are important for optoelectronic properties including finite-temperature band-gap energy distributions  {and possible band-to-band transitions.}
We contrast the findings for CsPbBr$_3$ to the case of PbTe, which shares several important characteristics with prototypical HaPs but, due to its rocksalt structure, cannot exhibit any transversality, in order to demonstrate that absence of this feature is detrimental to optoelectronic properties.
Our results suggests that the peculiar lattice vibrational properties of HaPs might not just be a mere coincidence, but rather could potentially play an active role in determining the favorable optoelectronic properties of these materials.

\begin{figure}
    \centering
    \includegraphics[width=0.5\textwidth]{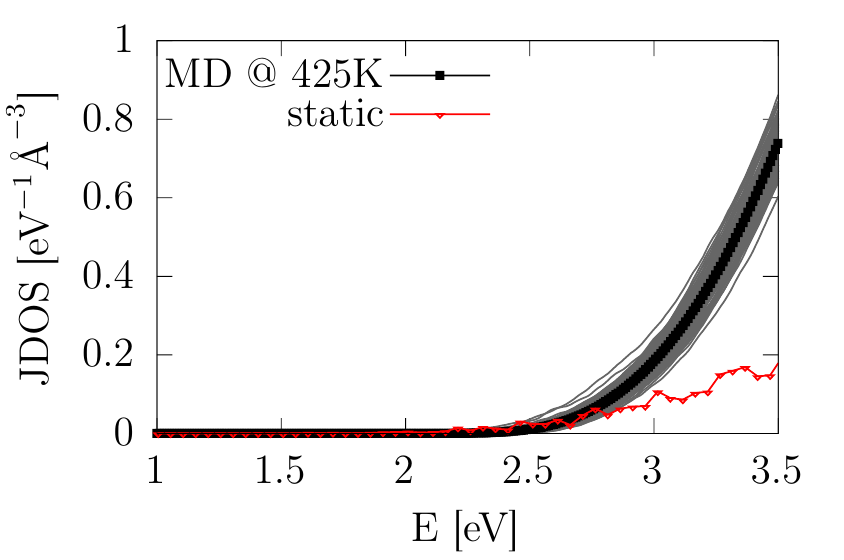}
    \caption{
    Joint-density of states (JDOS) of CsPbBr$_3$, calculated for a static cubic structure (red) and from molecular dynamics (MD) at $425~\mathrm{K}$ (black). Thin-black lines correspond to the JDOS of individual MD snapshots and the thick-dotted line to the their average.}
    \label{fig:JDOS1}
\end{figure}

We  {investigate} the protoypical HaP material CsPbBr$_3$, which is known to exhibit, on average, a cubic-symmetry structure of corner-sharing octahedra and void-filling A-site cations at temperatures above $\sim400$ K.\cite{Shunsuke1974,Rodova2003,Stoumpos2013}
 {We consider CsPbBr$_3$ in its cubic phase because this is the regime where the vibrational characteristics are particularly interesting, \textit{e.g.}, previous work found that a broad central peak emerges in the Raman spectrum of cubic CsPbBr$_3$ due to local polar fluctuations. }\cite{Yaffe2017}
Performing MD calculations based on density functional theory (DFT) at a temperature of $425~\mathrm{K}$ (see Methods section) allows for depicting the finite-temperature, real-time structure of cubic CsPbBr$_3$, including  {all vibrations excited at that temperature and} a description of the appearing vibrational anharmonicity to all orders. 
 {\textbf{Figure}{~\ref{fig:JDOS1}} shows the joint-density of states (JDOS) of CsPbBr$_3$, calculated either for a static cubic structure or for snapshots of instantaneous structures that were recorded along the MD trajectory at $425~\mathrm{K}$.
The JDOS connects the electronic structure of a system to its optical absorption, because it quantifies the possible band-to-band transitions in the single-particle band-structure assuming constant transition matrix elements.}\cite{Pankove1975,Yu2010}
 {Clearly, the two scenarios yield vastly different results: first, the onset of the JDOS on the energy axis differs by $\approx 0.5~\mathrm{eV}$ and, second, the JDOS rises much more rapidly in the $425~\mathrm{K}$ data.
The first finding is related to the fact that the band-gap energy of the averaged, cubic-symmetry structure calculated in a static DFT calculation differs substantially (by $\approx0.7~\mathrm{eV}$) from the average band-gap energy calculated as a time-average of the MD calculation at $425~\mathrm{K}$, as has been discussed in previous work.
The second finding, {\textit{i.e.}}, that the JDOS rises more rapidly in the $425~\mathrm{K}$ MD than in the static DFT calculation, is particularly interesting because previous work has highlighted that already static HaPs feature a JDOS that is steeper close to the absorption edge when compared to other semiconductors.}\cite{Yin2014,Davies2018a,Wuttig2021a}
 {Here, we find that the JDOS is intensified in the MD at $425~\mathrm{K}$ compared to the result of the static calculation.
Taken together, these findings demonstrate that vibrational features occurring in cubic CsPbBr$_3$ at $425~\mathrm{K}$ strongly impact the JDOS, and we will therefore investigate which patterns in the finite-temperature atomic dynamics are responsible for it.}

\begin{figure*}
    \centering
    \includegraphics[width=1.0\textwidth]{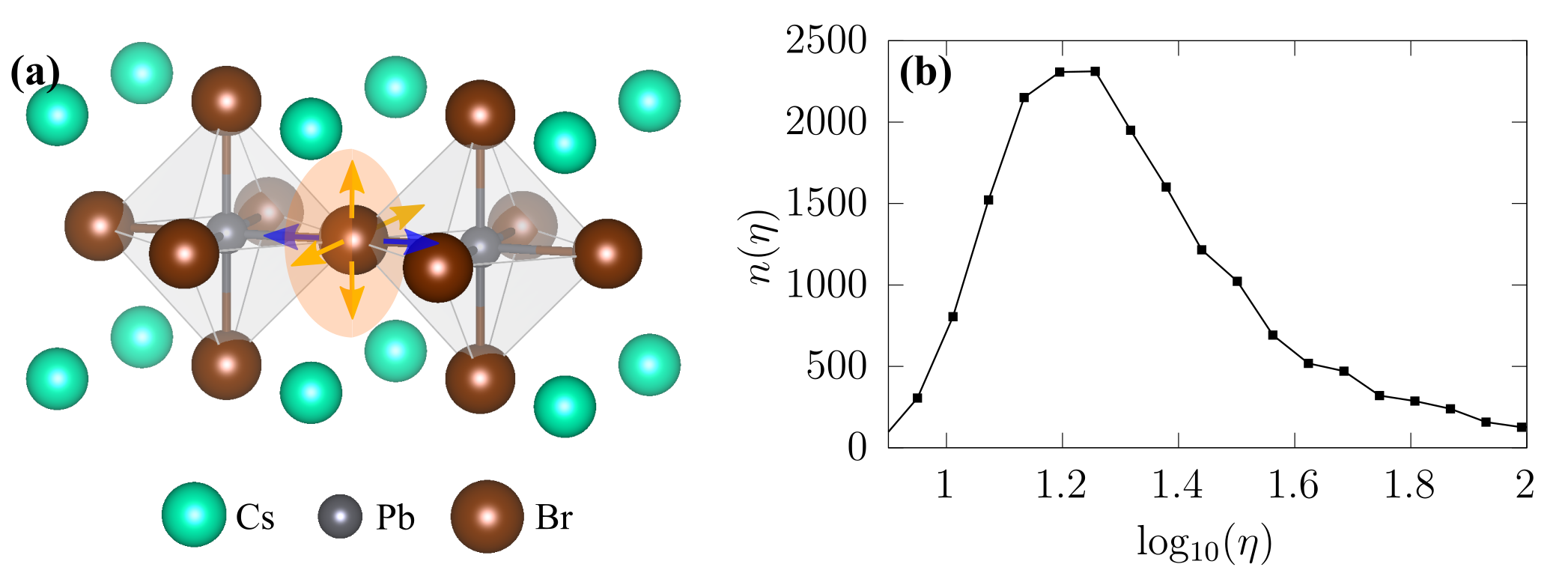}
    \caption{\textbf{(a)} Schematic structural representation of cubic CsPbBr$_3$ and the displacements of Br ions at finite-temperature: {longitudinal} components occur in parallel to the Pb-Br-Pb bonding axis (blue arrows) and {transversal} components perpendicular to it (yellow arrows expanding the light-orange plane).
    \textbf{(b)} Histogram of transversality, $\eta$, as defined in Eq.~\ref{eq:transversality}, calculated along the MD trajectory at 425~K. Note that this is shown as a semi-log plot.
    }
    \label{fig1}
\end{figure*}

Since electronic states close to the band edges of HaPs in general, and CsPbBr$_3$ in particular, 
are formed by the lead and halide atomic orbitals, we focus on the dynamics of the Pb-Br-Pb network and motions of halides therein, and stipulate to describe the finite-temperature Br displacements  {occurring in the MD} by two directional components: either  {Br atoms move} along the Pb-Br-Pb bond axis (blue line in \textbf{Figure}~\ref{fig1}a) or on the 2D plane that is perpendicular to that axis (light-orange plane in Figure~\ref{fig1}a). 
Thus, any Br displacement occurring at finite temperature is characterized by its \textit{longitudinal} (along the Pb-Br-Pb bond axis) and \textit{transversal} components (perpendicular to the Pb-Br-Pb bond axis).

For each structure occurring along the MD trajectory we hence define a transversality, $\eta$, as follows
\begin{equation}
    \eta = \frac{1}{N_{\mathrm{Br}}} \sum_{i=1}^{N_{\mathrm{Br}}} \frac{d_{i}^{\mathrm{transv}}}{d_{i}^{\mathrm{longi}}} \ ,
    \label{eq:transversality}
\end{equation}
where $N_{\mathrm{Br}}$ is the number of Br atoms in the supercell and $d_{i}^{\mathrm{transv/longi}}$ are the norms of the transversal~/ longitudinal displacement components of the $i$th Br ion.
We present additional means of comparing the occurrence of transversal and longitudinal displacements in the Supporting Information (SI).
Calculating the histogram of $\eta$ along the MD trajectory (see {Figure}~\ref{fig1}b) reveals that CsPbBr$_3$ in its cubic structure features large transversal Br displacements: the histogram peaks at $\eta\approx20$, with a significant tail at high values of $\eta$, up to $\eta\approx100$, and quickly decreases to zero when approaching $\eta=1$.
This indicates that the movements of Br ions are very much favored on the 2D-planes perpendicular to Pb-Br-Pb bond axis compared to movements along it, which suggests a peculiar {directionality} in the finite-temperature dynamics of CsPbBr$_3$ despite the fact that, {on average}, it exhibits a cubic symmetry.
Interestingly, the appearance of transversality is closely connected to the rotations of the 
PbBr$_6$ octahedra (see SI),which are a known generic feature of the finite-temperature dynamics of HaPs.\cite{Smith2015,Young2016,Ghosh2017,Klarbring2019} 
In the SI, we also show that the transversal Br displacements are concurrent with lattice-vibrational anharmoncity in CsPbBr$_3$.

\begin{figure}
    \centering
    \includegraphics[width=0.5\textwidth]{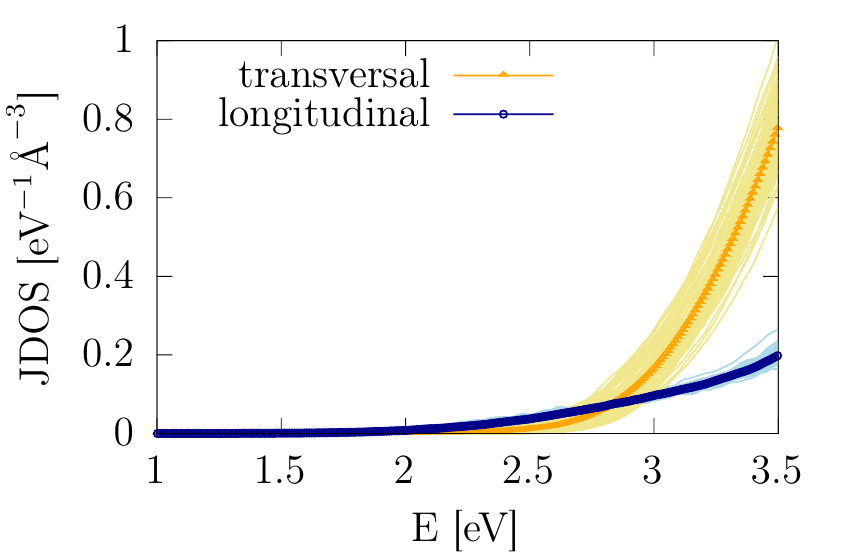}
    \caption{ {
    MD-calculated JDOS of CsPbBr$_3$ at $425~\mathrm{K}$ corresponding to Br motions that occur either along longitudinal (blue) or transversal (yellow) directions. Thin lines correspond to data of individual snapshots and thick-dotted lines to their average.
    }}
    \label{fig:JDOS2}
\end{figure}

The large degree of transversality in the Br ion dynamics signifies an enormous dynamic structural flexibility of HaPs because  {Br ions} that participate in covalent bonding with  {two adjacent Pb ions} are actually favorably displaced \textit{away from the  {Pb-Br-Pb} bonding axis}. 
The occurrence of this property motivates an investigation into its origin and implications for other important characteristics of CsPbBr$_3$.
 {We therefore calculate the JDOS again from MD of CsPbBr$_3$ at $425~\mathrm{K}$, but now separately for longitudinal and transversal displacements of Br atoms (see \textbf{Figure}~{\ref{fig:JDOS2}}).
Interestingly, this procedure finds results that essentially mimic what we have reported above for the JDOS in case of the static lattice and at $425~\mathrm{K}$.
Most importantly, the transversal version of the JDOS rises much faster than the longitudinal one, such that there is a clear connection between the occurrence of transversality and a rapid increase in JDOS.
This finding illustrates a remarkable aspect in how the dynamic structural flexibility of HaPs determines their optoelectronic properties, namely such that transversal Br motion intensifies the possible band-to-band transitions in CsPbBr$_3$.}

\begin{figure*}
    \centering
    \includegraphics[width=1.0\textwidth]{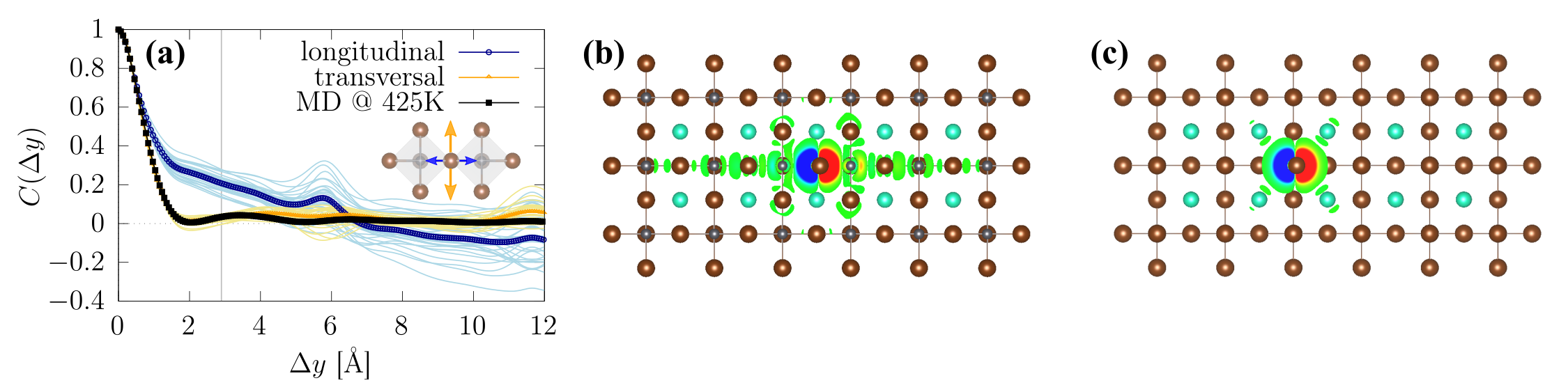}
    \caption{
    \textbf{(a)} Autocorrelation function of the disorder potential, $C(\Delta y)$, that is induced by the Br motions along {longitudinal} (blue) and {transversal} (yellow) directions in CsPbBr$_3$ at $425~\mathrm{K}$.
    The thin lines correspond to individual snapshots, $C_i(\Delta y)$ in Eq.~(\ref{eq:correlation}), and the thick-dotted lines to their average.
    $C(\Delta y)$ according to the full set of displacements occurring in the  $425~\mathrm{K}$ trajectory is also shown (black line).
    The inset shows two adjacent PbBr$_6$ octahedra and depicts the Br displacement directions, and the vertical gray line indicates the nominal Pb-Br bonding distance.
    Isosurface representation of the calculated charge-density response to a Br displacement along \textbf{(b)} {longitudinal} direction and \textbf{(c)} {transversal} direction in a supercell of CsPbBr$_3$. Isosurfaces are shown for changes above $8.1\times10^{-3}~e\mathrm{\AA^{-3}}$ ($e$ being the electron charge).
    }
    \label{fig3}
\end{figure*}

 {To investigate the origin of dynamic structural flexibility and large degree of transversality in CsPbBr$_3$,} we analyze the disorder potential generated by the finite-temperature ion dynamics.
Specifically, we compute the autocorrelation function of the disorder potential, $C(\Delta y)$ (see Methods section), for which we have previously found  that it is dynamically shortened and approaches a value of zero on length scales on the order of one Pb-X bond.\cite{Gehrmann2019}
\textbf{Figure}~\ref{fig3}a shows $C(\Delta y)$ separately for the longitudinal and transversal components of Br displacements in CsPbBr$_3$ at 425 K (see Methods section), in addition to the result from a full MD calculation.
While the longitudinal components alone would lead to a long-ranged response, clearly the transversal Br displacements are responsible for the short-ranged nature of the disorder potential in CsPbBr$_3$ at finite temperature.
Such a short-ranged disorder potential is known to be concurrent with narrower band-edge distributions that imply small values for the Urbach energy.\cite{Gehrmann2019,Greeff1995} 
 {Keeping in mind that other effects are of course also potentially important for determing the Urbach energy of HaPs,}\cite{Savill2021a}
 {the finding shows that the transversality in the Br dynamics is an important factor that leads to a rapid rise of the JDOS and sharp optical absorption.}

The reason for the transversality in Br displacements and the short-ranged nature of the disorder potential can be found in the very peculiar way the HaP system responds to changes in its atomic configurations.
To illustrate this, we calculated the charge-density response to specific atomic displacements in an otherwise static supercell of CsPbBr$_3$ using DFT (see Methods section for details).
Displacing a Br ion longitudinally leads to a response in the charge density that is clearly \textit{long-ranged} and spans several unit cells (Figure~\ref{fig3}b).
In sharp contrast to this scenario, displacing a Br ion transversely leads to a response in the charge density that is clearly \textit{short-ranged} and essentially confined to a single unit cell (Figure~\ref{fig3}c).
This qualitative difference in the charge-density response comparing transversal and longitudinal Br displacements is related to the resonant bonding mechanism\cite{Krebs1955,Krebs1956} that has been extensively discussed in the literature on HaPs\cite{Weber1978,Zhu2019, Gehrmann2019} and other materials:\cite{Lucovsky1973,Shportko2008,Lee2014,Yue2018} 
along the Pb-Br-Pb bond axes, Br-4p$_\mathrm{x}$ and Pb-6s $\mathrm{\sigma}$-interactions give rise to the valence band and a network of resonant bonds.\cite{Kim2015,Goesten2018} 
Displacing an atom along the direction of orbital hybridization leads to changes in the charge density that ``resonate'' throughout the network: when the symmetry of the network is broken by such a displacement, degeneracies of orbital configurations are lifted, the orbital occupations thus modulated, and a long-ranged response in the charge density must follow  (\textit{cf.} Figure~\ref{fig3}b).
In contrast, any transversal displacement cannot influence the resonant bonding network, since by its definition the transversal plane is orthogonal to the direction of the network, as are the Br-p$_\mathrm{y}$ and Br-p$_\mathrm{z}$ states which do not participate in the resonant bonding.
Therefore, the transversal Br displacements essentially do not perturb the resonant network and, hence, only lead to short-ranged changes in the charge density (\textit{cf.} Figure~\ref{fig3}c).
It is for this reason that transversal Br displacements are highly favored energetically (see SI) over the longitudinal ones and, hence, occur with a much larger likelihood.

\begin{figure*}
    \centering
    \includegraphics[width=0.8\textwidth]{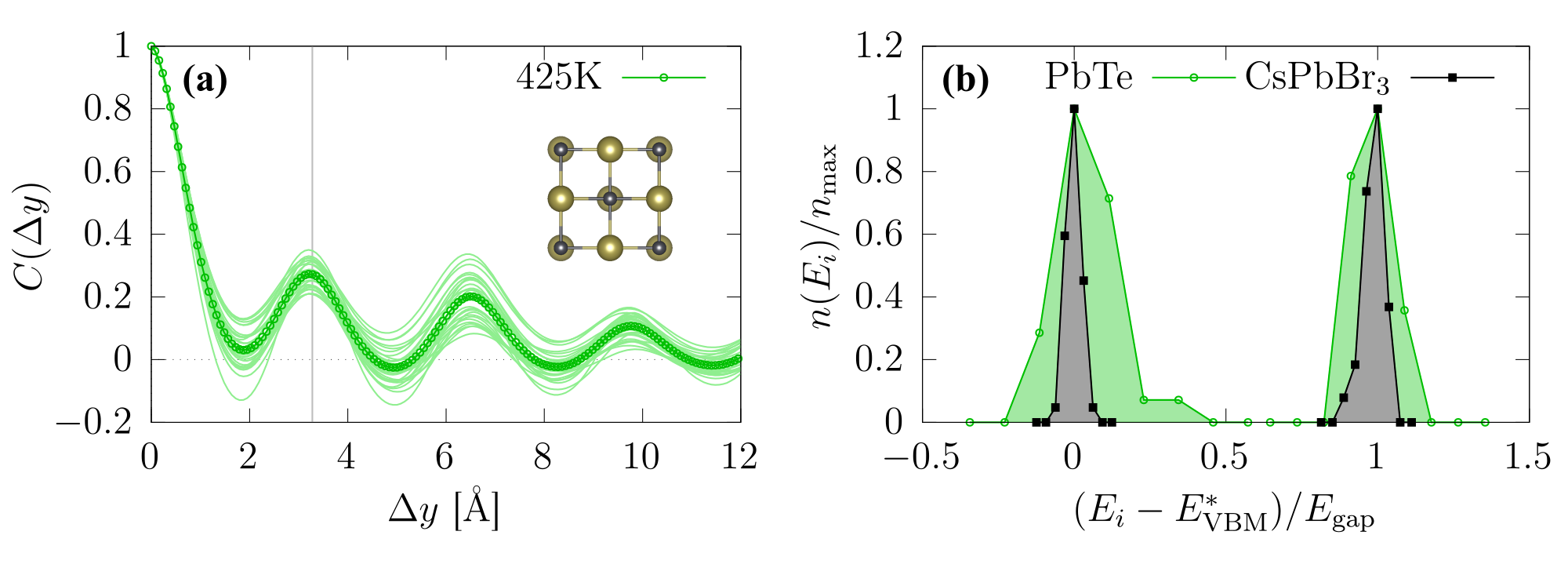}
    \caption{\textbf{(a)} $C(\Delta y)$ for PbTe material calculated at $425~\mathrm{K}$; the inset depicts a structural representation of this system and the gray vertical line indicates the nominal Pb-Te bonding distance. Panel \textbf{(b)} shows normalized histograms of the band-edge energies calculated along the MD trajectories of CsPbBr$_3$ (black curve) and PbTe (green curve) at $425~\mathrm{K}$. 
    The histograms were centered to the VBM/CBM energy with highest occurrence, $E_{\mathrm{VBM/CBM}}^{*}$,
    in order to allow for a direct comparison of the band-edge distributions of CsPbBr$_3$ and PbTe (see Methods section).
   }
    \label{fig4}
\end{figure*}

Our findings suggest that the dynamic structural flexibility of HaPs, such as CsPbBr$_3$ with its apparent degree of transversality, may play an active role in determining the favorable optoelectronic properties of HaPs, because a short-ranged disorder potential implies sharper optical absorption onsets.
To test this hypothesis, we consider PbTe as a counter-example for a material sharing many key properties with CsPbBr$_3$ (see SI for detailed discussion): bonding properties including the anti-bonding nature of the VBM formed by $\mathrm{\sigma}$ interactions of anion p and cation s orbitals, the cross-gap hybridization of the anion p orbitals, a stereochemically active lone pair of electrons\cite{Wei1997,Waghmare2003,Brod2020} as well as vibrational anharmonicities\cite{Delaire2011,Li2014a} 
and, interestingly, a similar type of resonant bonding mechanism\cite{Lee2014}.
Importantly, however, given its rocksalt structure (see inset in \textbf{Figure}~\ref{fig4}a), from symmetry considerations PbTe cannot exhibit any degree of transversality.
As a consequence of the absence of transversality, any atomic displacement must perturb the resonant network to a certain degree, which leads to longer-ranged changes in the charge-density response to atomic displacements, as we explicitly show in the SI.
In full agreement with the expectations borne from this reasoning, we find features in $C(\Delta y)$ for PbTe (see {Figure}~\ref{fig4}a) reaching beyond several unit cells when we perform MD at 425 K, which is in sharp contrast to the ultra short-ranged correlations found in CsPbBr$_3$ (\textit{cf.} Figure~\ref{fig3}a); note that the oscillatory features can be explained by {thermal noise that is due to the atomic displacements modulating the crystal periodic potential} (see SI). 
Most importantly, comparing the conduction and valence band histograms of PbTe and CsPbBr$_3$ (see Figure~\ref{fig4}b) shows that the absence of transversality leads to much broader distributions, which implies a larger amount of thermally-induced tail-states and energetic disorder and, thus, less sharp optical absorption edges and larger Urbach energies.
We therefore speculate that the much larger Urbach energy of PbTe\cite{Wang2008b} compared to CsPbBr$_3$\cite{Rakita2016} could in part be due to its more long-ranged disorder correlations that we have demonstrated here to be a direct consequence of the absence of transversality, keeping in mind that other extrinsic factors may impact the Urbach energy of HaPs too.\cite{Savill2021a}
Altogether, these findings demonstrate that the dynamic structural flexibility of CsPbBr$_3$ that we characterized here by the large degree of transversality in the Br displacements, potentially plays an important role in establishing the favorable optoelectronic properties of HaPs.

Finally, we would like to put our work in context of recent findings in the literature.
Several recent studies reported an important role of specific vibrational modes that are very similar or at least closely related to the transversal modes we highlight here: \textit{e.g.}, the two-dimensional overdamped fluctuations recently found in CsPbBr$_3$\cite{Lanigan-Atkins2021} as well as the X-Pb-X scissoring mode.\cite{Bird2021}, both of which have also been discussed to imply certain dynamic modulations to the band gap.\cite{Lanigan-Atkins2021,Bird2021}
Another case in point that these modes are relevant for the optoelectronic properties of HaPs is that their counterparts, \textit{i.e.}, the longitudinal modes, were previously found to be involved in so-called cage vibrations that were discussed to be the main finite-temperature contributors to the Urbach energy.\cite{Ledinsky2019a} 
This assessment is in full agreement with the expectations borne from our findings that disorder correlations associated with longitudinal displacements would be longer ranged (\textit{cf.} Figure~\ref{fig3}).
Therefore, we believe that our findings motivate a wider investigation of how the dynamic structural flexibility in HaPs and related materials could be exploited towards design of materials with improved properties.

In summary, we investigated the  {effect of the dynamic structural flexibility of HaPs on their optoelectronic properties} by means of first-principles MD for the prototypical model system CsPbBr$_3$ in the cubic phase.
We found that the vibrations in the HaP lattice feature a distinctive property of transversality, which describes the large displacements of halide ions along directions that are orthogonal to the Pb-Br-Pb bonding axis. 
It was demonstrated that this property leads to  {a rapid rise of the JDOS that is concurrent with} a shortening of disorder correlations, which implies a sharpening of the distribution of band edges, a reduction of the number of thermally induced tail-states and, by extension, a  {sharper optical absorption profile}.
 {Since the related Urbach tail determines the losses of open-circuit voltage in the radiative limit,{\cite{Bisquert2021}} this finding is potentially relevant for the design of solar materials.}
We have also contrasted these findings to the case of PbTe, a similarly anharmonic material also exhibiting resonant bonding and a similar electronic structure, but lacking the dynamic structural flexibility of the perovskite lattice as expressed in transversality.
Our findings establish the important link between dynamic structural flexibility of HaPs with their favorable optoelectronic properties. 
This suggests that the unusual lattice vibrational properties of HaPs might not just be a mere coincidence and could be used to guide the material design of new compounds with similarly favorable or even enhanced optoelectronic properties.

\section{Methods}
\threesubsection{Static DFT Calculations}
DFT calculations were performed with the Vienna Ab-initio Simulation Package (VASP) code,\cite{Kresse1996}
using the projector-augmented wave (PAW) method\cite{Kresse1999} applying the "normal" version of the code-supplied PAW potentials unless otherwise noted. 
Exchange-correlation interactions were described using the Perdew-Burke-Ernzerhof (PBE) form of the generalized gradient approximation,\cite{Perdew1996a}
including corrections for dispersive interactions according to the Tkatchenko-Scheffler scheme.\cite{Tkatchenko2009}
This setup has been shown to allow for an accurate description of the finite-temperature structural dynamics of HaPs.\cite{Beck2019}
The ionic and lattice degrees-of-freedom of CsPbBr$_3$ were optimized, starting from a cubic structure, resulting in a final lattice constant of $5.81~\mathrm{\AA}$.
For PbTe, we used a conventional rocksalt unit cell (8 atoms), obtained from the Crystallography Open Database,\cite{Graulis2009} (COD-ID: 9008696)\cite{Wyckoff1963} and relaxed it to a lattice constant of $6.56~\mathrm{\AA}$.
For structural relaxation and static DFT calculations, an energy convergence threshold of $10^{-8}~\mathrm{eV}$, a $\mathrm{\Gamma}$-centered $k$- grid of $6\times6\times6$ and a plane-wave cutoff energy of $500~\mathrm{eV}$ (CsPbBr$_3$) and $375~\mathrm{eV}$ (PbTe) were used.
The geometries of CsPbbr$_3$ and PbTe were optimized until residual forces were below $10^{-3}~\mathrm{eV\AA^{-1}}$. 

\threesubsection{First-Principles MD Calculations}
DFT-MD simulations were performed on a $4\times4\times2$ (160 atoms) supercell of CsPbBr$_3$ and a $3\times3\times2$ (144 atoms) of PbTe, using the canonical (\textit{NVT}) ensemble at $T=425~\mathrm{K}$ with a Nos\'{e}-Hoover thermostat as implemented in the VASP code.\cite{Kresse1994} 
The simulation timestep was set to $8~\mathrm{fs}$ for CsPbBr$_3$ and $10~\mathrm{fs}$ for PbTe.
For the sake of efficiency, different settings for the self-consistent calculations in each ionic step were employed:
the "GW" PAW potentials were used, together with plane-wave cutoffs of $250~\mathrm{eV}$ (CsPbBr$_3$) and $240~\mathrm{eV}$ (PbTe), a single $k$-point and an energy convergence threshold of $10^{-6}~\mathrm{eV}$.
The system was equilibrated for $5~\mathrm{ps}$ and a subsequent trajectory of $150~\mathrm{ps}$ was used for analysis.

\threesubsection{JDOS, Disorder Potentials and Band-Edge Histograms}
JDOS, disorder potential and band-edge histograms were calculated from snapshots of instantaneous configurations along the MD trajectory.
The charge density and electrostatic potentials were then computed using a $1\times1\times2$ $k$-grid for CsPbBr$_3$ and $2\times2\times3$ $k$-grid for PbTe, preserving the density of $k$-points in the supercell as compared with the unit-cell calculations.
 {To obtain the JDOS, we subsequently performed non-selfconsistent calculations on a finer, $3\times3\times6$ $k$-grid, for 90 snapshots that were evenly distributed along the trajectory and had filtered out those structures whose geometries resulted in band-gap energy lower than 80\% of the average value.
The JDOS was then calculated from the eigenvalues, $\varepsilon_i(\mathbf{k})$, as:}\cite{Wang2021}

\begin{equation}\label{eq:jdos}
    j(E) = 2 \sum_{v,c,\mathbf{k}}{\frac{w_\mathbf{k}}{\sigma \sqrt{2\pi}}\mathrm{e}^{ -\frac{(\varepsilon_c(\mathbf{k})-\varepsilon_v(\mathbf{k})-E)^2}{2\sigma^2}}}
\end{equation}
 {where the $v$ and $c$ indices signify valence and conduction bands, respectively, $\mathbf{k}$ the $k$-points with their respective weighting $w_\mathbf{k}$, and $\sigma$ the spectral broadening that we have set to 20~meV.}

The autocorrelation, $C(\Delta y)$, was calculated --- as detailed in our previous work --- as the average of the instantaneous autocorrelations:\cite{Gehrmann2019}

\begin{equation} \label{eq:correlation}
     C_i(\Delta y) = \frac{\langle \Delta V_{i}(y+\Delta y)\cdot \Delta V_{i}(y) \rangle}{\langle \Delta V_{i}(y)\cdot \Delta V_{i}(y) \rangle}
    \end{equation}
where $\Delta V_{i}$ is the $xz$-averaged change in the electrostatic potential of configuration $i$ with respect to the average potential along the trajectory,
and $C(\Delta y)$ is averaged over $N=30$ instantaneous configurations.
To separate the effect of longitudinal and transversal Br displacements in the calculations of $C(\Delta y)$, we modified the MD trajectory as follows: for each Br, we separated the $x$, $y$ and $z$ cartesian components of a given displacement into longitudinal and transversal contributions (\textit{cf.} Figure~\ref{fig1}a), setting either one of them as well as Cs and Pb displacements to zero. 
The CBM an VBM distributions in Figure \ref{fig4}b were calculated using the respective eigenvalues obtained from DFT calculations performed for 90
configurations along the MD trajectory. 
These histograms were centered to the VBM/CBM energy with highest occurrence, $E_{\mathrm{VBM/CBM}}^{*}$, using $E_{\mathrm{gap}}^{*}=E_{\mathrm{CBM}}^{*}-E_{\mathrm{VBM}}^{*}$ normalizing to the respective count, that is $n_{\mathrm{max}}=n(E_{\mathrm{VBM/CBM}}^{*})$.
The normalization allows for a direct comparison of the band-edge distributions of CsPbBr$_3$ and PbTe despite their different values of the band gap.

\medskip
\textbf{Supporting Information} \par 
Supporting Information is available from the Wiley Online Library or from the author.

\medskip
\textbf{Acknowledgements} \par 
The authors thank Laura Herz (University of Oxford) and Omer Yaffe (Weizmann Institute of Science) for fruitful discussions.
Funding provided by the Alexander von Humboldt-Foundation in the framework of the Sofja Kovalevskaja Award, endowed by the German Federal Ministry of Education and Research, by
the Deutsche Forschungsgemeinschaft (DFG, German Research Foundation) via SPP2196 Priority Program (project-ID: 424709454) and via Germany's Excellence Strategy - EXC 2089/1-390776260, and by the Technical University of Munich - Institute for Advanced Study, funded by the German Excellence Initiative and the European Union Seventh Framework Programme under Grant Agreement No. 291763, are gratefully acknowledged.
The authors further acknowledge the Gauss Centre for Supercomputing e.V. for funding this project by providing computing time through the John von Neumann Institute for Computing on the GCS Supercomputer JUWELS at J\"ulich Supercomputing Centre.

\medskip

\printbibliography


\end{document}